# Significance of Strain Rate in Severe Plastic Deformation on Steady-State Microstructure and Strength

Kaveh Edalati[1,*], Qing Wang[1], Nariman A. Enikeev[2,3], Laura-Jean Peters[4], Michael J. Zehetbauer[4] and Erhard Schafler[4]

[1] WPI, International Institute for Carbon-Neutral Energy Research (WPI-I2CNER), Kyushu University, Fukuoka 819-0395, Japan
[2] Ufa State Aviation Technical University (USATU), K. Marx 12, 450008 Ufa, Russia
[3] Center for Design of Functional Materials, Bashkir State University, 450076 Ufa, Russia
[4] Faculty of Physics, University of Vienna, Boltzmanngasse 5, A-1090 Wien, Austria

The microstructure and mechanical properties of materials saturate to steady states after severe plastic deformation (SPD). Despite the well-known effect of temperature on the steady-state microstructure, there is no general agreement on the significance of strain rate and the applicability of the Zener-Hollomon parameter in this regard. In this study, several pure metals (aluminum, copper, titanium, and iron) and a Cu-30Zn (wt%) brass alloy have been processed by a high-speed high-pressure torsion (HPT) equipment with controllable rotation speeds in the range of 0.06 to 60 rpm. It is found that crystallite/grain size, dislocation density, microhardness and shear stress at the steady state are reasonably rate-independent for the von Mises strain rates in the range of 0.004 to 20 s$^{-1}$. Because both rates of grain refinement and of dynamic recrystallization are proportional to the strain rate, it is suggested that their balance, which determines the steady state, is rate-independent.

*Corresponding author (E-mail: kaveh.edalati@kyudai.jp; Tel: +81-92-802-6744)

**Introduction**

Severe plastic deformation (SPD) methods are efficient in achieving ultrafine-grained (UFG) microstructures in a wide range of metallic and non-metallic materials [1]. The microstructure and resultant mechanical and functional properties significantly change at the early stages of straining, but they finally saturate to the steady state at large strains particularly in single-phase materials [1]. The occurrence of a steady state, which was first recognized by Bridgman in the 1930s [2], has been attributed to a balance between the rate of grain refinement and defect generation on the one hand, and the rate of dynamic recovery, recrystallization and grain boundary migration on the other hand [3-5].

Investigation of the parameters influencing the steady-state grain size has been of significant interest for the past two decades because one target in SPD processing has been introducing new strategies for further reduction of the final grain size [4-6]. These investigations sometimes reported contradicting results. For example, while several studies suggested that stacking fault energy is the most important factor in determining the steady-state grain size [7,8], some others suggested that the final grain size is reasonably independent of stacking fault energy [9,10]. Most of the studies, however, suggested that homologous temperature and atomic diffusion are important factors in determining the steady-state grain size [3-6]. Despite a general agreement on the effect of temperature on the steady-state grain size, the significance of strain rate has not been well clarified so far.

In conventional metal forming, it is generally accepted that strain rate and temperature affect the microstructural and strength evolution through the Zener-Hollomon parameter, $Z = \dot{\varepsilon} \exp^{-Q/RT}$ ($\dot{\varepsilon}$: strain rate, $Q$: activation energy for the operative thermally activated process, $R$: gas constant, $T$: absolute temperature) [11]. In SPD processing, some researchers suggested that higher strain rates can reduce grain size, particularly when the homologous temperature is low [12-16], while others did not find any rate-dependent changes at the steady state [17]. Several studies showed that the strain rate can influence the steady-state microstructure through a change in the deformation mechanism (e.g., dislocation activity to twinning) [18] or phase transformation [19], while other researchers reported only a small change in the microstructure even at extremely large strain rates [20]. Since a high strain rate can enhance the rate of both grain fragmentation [4,11] and dynamic recrystallization [21,22] phenomena, its influence on the steady-state grain size, where there is a balance between these two phenomena, still needs clarification.

In this study, high-pressure torsion (HPT), as an SPD method [1,2], is employed and the effect of von Mises strain rate in the range of 0.004 s$^{-1}$ to 20 s$^{-1}$ on the steady-state microstructure, hardness and shear strength is studied for five model materials (aluminum, copper, titanium, iron and brass). The results confirm that the steady state is independent of strain rate within the detection limits of employed characterization methods.

**Experimental Procedures**

Discs of Al 1050 (99.5%), Cu (99.99%), Ti (99.9%), Fe (99.96%) and Cu - 30 wt% Zn with 10 mm diameter and 0.8 mm thickness were annealed in an argon atmosphere for 1 h at 773, 873, 1073, 1273 and 793 K, respectively, and processed by a high-speed HPT machine. As shown in Fig.. 1a and 1b, the machine was equipped with a hydraulic press, an electric motor and a control unit, and it had two anvils made of a composite of tungsten carbide and 11 wt% cobalt. There was a 10 mm diameter flat-bottomed hole with 0.25 mm depth and 0.1 mm roughness on the center of each anvil to account for full torsion deformation of disc samples. The HPT process was conducted for 15 turns under a pressure of 6 GPa for Al, Cu, Fe and Cu-Zn and under 2 GPa for Ti to avoid the formation of the high-pressure ω-Ti phase [19]. The rotation speed of HPT anvils was either 0.06, 1, 8 or 60 rotations per minute (rpm). The maximum temperature of anvils, which was measured during the process using a thermocouple located at 10 mm away from the disc in the upper anvil or by an optical thermometer, was below 323 K. The Fe discs processed for 60 rpm were excluded from the experiments because they stack to the anvils, which made the reliability of their processing unclear. The surfaces of disc samples before and after HPT processing with 1 rpm are shown in Fig. 1c. The roughness of the sample surfaces was similar to the roughness of flat-bottomed holes on the anvils, which was introduced for avoiding slippage of samples during the HPT process. After the HPT process, the discs were first ground using sandpapers to remove the roughened surface layers and then metallographically polished for examination by different methods of microstructural and mechanical property characterization.

First, the discs were polished to mirror-like surfaces and the Vickers hardness on the upper surface of the discs was measured in four different radial directions at 0.15, 1, 2, 3 and 4 mm away from the disc center using the indentation loads of 2 N for Al, 3 N for Cu and 5 N for Ti, Fe and Cu-Zn and the indentation time of 15 s.

Second, in addition to the *ex-situ* evaluations of strength by hardness measurement, the strength of Fe was examined *in situ* by torque measurements during HPT under a pressure of 4 GPa for 2 turns with rotation speeds of 0.1, 0.6, 1 and 6 rpm. The shear stress was estimated as $\tau = \frac{3q}{2\pi R^3}$ where $q$ is the measured torque, and $R$ is the radius of the disc (see, e.g. [14]).

Third, the polished discs were examined by X-ray diffraction (XRD) with the Rigaku SmartLab diffractometer using Cu Kα radiation. The XRD profiles were evaluated by the Rietveld refinement by the MAUD software to determine the microstrain and crystallite size [23]. The crystallite size is the average dimension of defect-free crystalline areas (domains) that scatter X-rays coherently [24]. It was shown that crystallite size in severely deformed metals is equivalent to subgrain or dislocation cell size [24]. Since the strength of deformed and severely deformed metallic materials is mainly determined by dislocation cell sizes [25,26], the crystallite size is an important parameter to clarify the significance of strain rate dependence in this study. For the Rietveld refinement, instrumental broadening was considered by measuring the silicon standard sample. Moreover, the dislocation density was determined using the Williamson-Smallman method [27] for the materials with cubic symmetry and using the technique proposed in [28] for <*a*>-type dislocations in Ti with hexagonal symmetry.

Fourth, the microstructure of samples was additionally examined by transmission electron microscopy (TEM). Discs with a 3 mm diameter were cut from the edge of HPT-processed samples using electric discharge machining and mechanically thinned to a thickness of 0.10-0.15 mm using sandpapers. The 3 mm discs were further thinned to foils with electron transparency using a twin-jet electro-polisher for Al (solution: 15 vol% $HClO_4$ + 15 vol% $C_3H_5(OH)_3$ + 70 vol% $CH_3OH$, voltage: 13 V, temperature: 263 K), Cu (solution: 15 vol% $HNO_3$ + 15 vol% $C_3H_5(OH)_3$ + 70 vol% $CH_3OH$, voltage: 10 V, temperature: 263 K), Ti (5 vol% $HClO_4$ + 25 vol% $C_3H_3(CH_2)_2CH_2OH$ + 70 vol% $CH_3OH$, voltage: 13 V, temperature: 263 K), Fe (90 vol% $CH_3COOH$ + 10 vol% $HClO_4$ voltage: 14 V, temperature: 300 K) and Cu-Zn (solution: 20 vol% $HNO_3$ + 80 vol% $CH_3OH$, voltage: 20 V, temperature: 263 K). The thin foils were examined by TEM under an acceleration voltage of 200 keV or 300 keV using bright-field images, dark-field images and selected area electron diffraction (SAED) patterns.

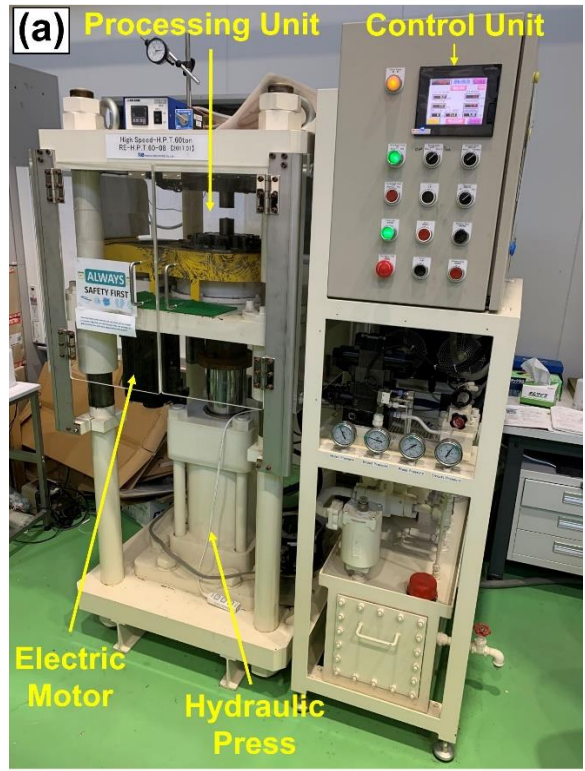
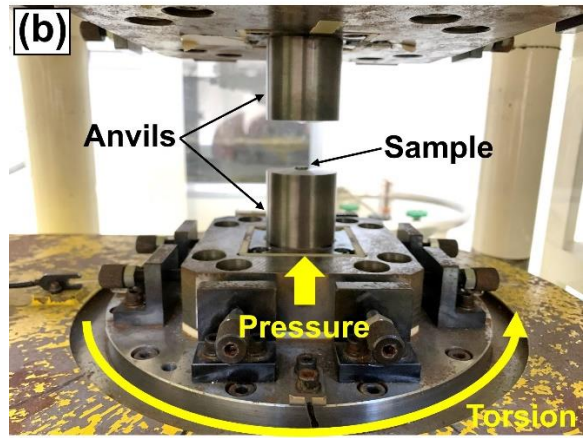
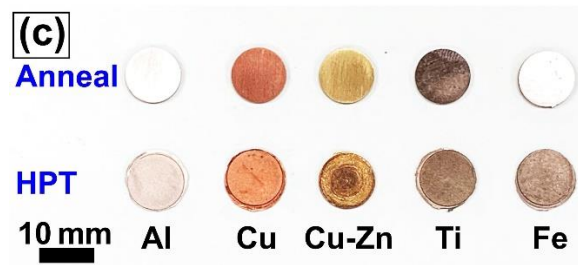

Figure 1. (a) Photograph of high-speed HPT machine. (b) Photograph of processing unit of HPT machine. (c) Surfaces of disc samples before and after HPT processing with 1 rpm.

**Results**

Fig. 2a shows the typical variation of hardness against distance from the disc center for Cu-Zn discs processed by HPT with different rotation speeds. The hardness values are at the steady state, except for the center of discs which show slightly lower hardness. The variation of hardness against the distance from the center for the four selected metals is like the one for Cu-Zn, although the magnitude of hardness depends on the material. The hardness values at different distances from the disc center excluding the center of discs are summarized in Fig. 1b as the steady-state hardness versus strain rate. The strain rate at each radial distance was estimated as von Mises shear strain ($\varepsilon = \gamma/\sqrt{3}$, where $\gamma = 2\pi rN/h$, $r$: radial distance, $N$: the number of turns and $h$: thickness [2]) divided by processing time ($t = N/\omega$, $\omega$: rotation speed). The steady-state hardness appears to be independent of strain rate within the range of 0.004 to 20 s$^{-1}$. The steady-state hardness is the lowest for Al and is higher for Cu, Cu-Zn, Ti and Fe, respectively, in good agreement with earlier publications [9,10].

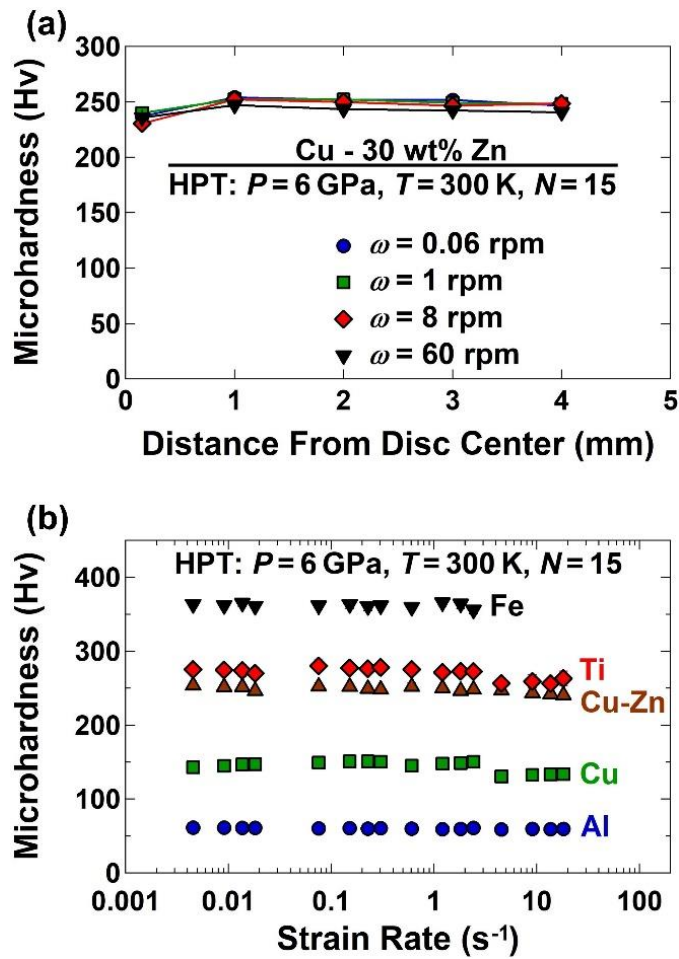

Figure 2. Effect of strain rate on steady-state hardness. Variations of hardness versus (a) distance from disc center and (b) strain rate for (a) Cu-Zn and (b) Al, Cu, Cu-Zn, Ti and Fe processed by HPT for 15 rotations with various rotation speeds.

Fig. 3 shows the variations of shear strength versus HPT rotations examined *in situ* by torque measurements for Fe processed by HPT with different rotation speeds. Despite some deviations in the stress-strain plots, the steady-state stresses at large stains appear to be independent of the rotation speed. Therefore, both *in-situ* torque measurements and *ex-situ* hardness measurements suggest that the flow stress is reasonably independent of strain rate, although an earlier study suggested a rate-dependent torque behavior for severely deformed Fe [14].

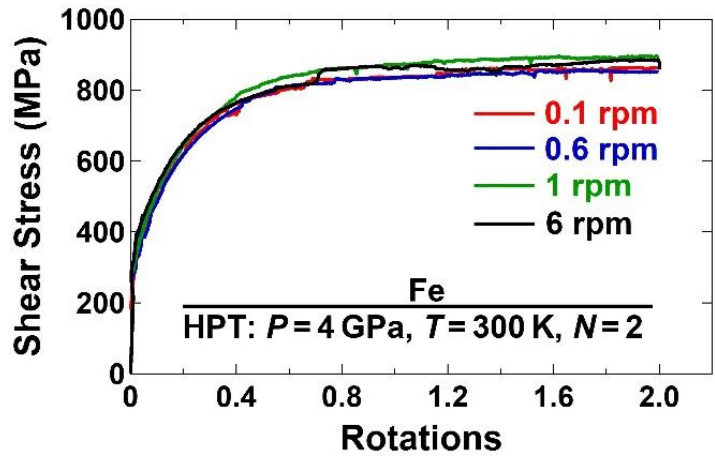

Figure 3. Effect of strain rate on steady-state shear stress. Shear stress versus the number of rotations, evaluated *in-situ* by torque measurement, for Fe processed by HPT by up to 2 rotations with various rotation speeds. Each curve represents the average of 4-6 measurements with an average standard deviation of 14 MPa (+/- 3 %) for shear stress.

Fig. 4a shows the typical XRD profiles for Cu-Zn processed by HPT with different rotation speeds. The material has an FCC structure and does not show any phase transformations by HPT processing. The XRD profiles for the four model metals also show the presence of single phases (FCC for Al and Cu, HCP for Ti and BCC for Fe) without the occurrence of any phase transformations after HPT processing. The XRD profiles for all materials were evaluated by the Rietveld refinement to determine the crystallite size and dislocation density, as shown in Fig. 4b and 4c, respectively. Both crystallite size and dislocation density are reasonably independent of strain rate, although the dislocation density slightly decreases, and the crystallite size slightly increases in some materials for a rotation speed of 60 rpm. These changes at 60 rpm can be due to the localized temperature rise during high-speed HPT [29,30].

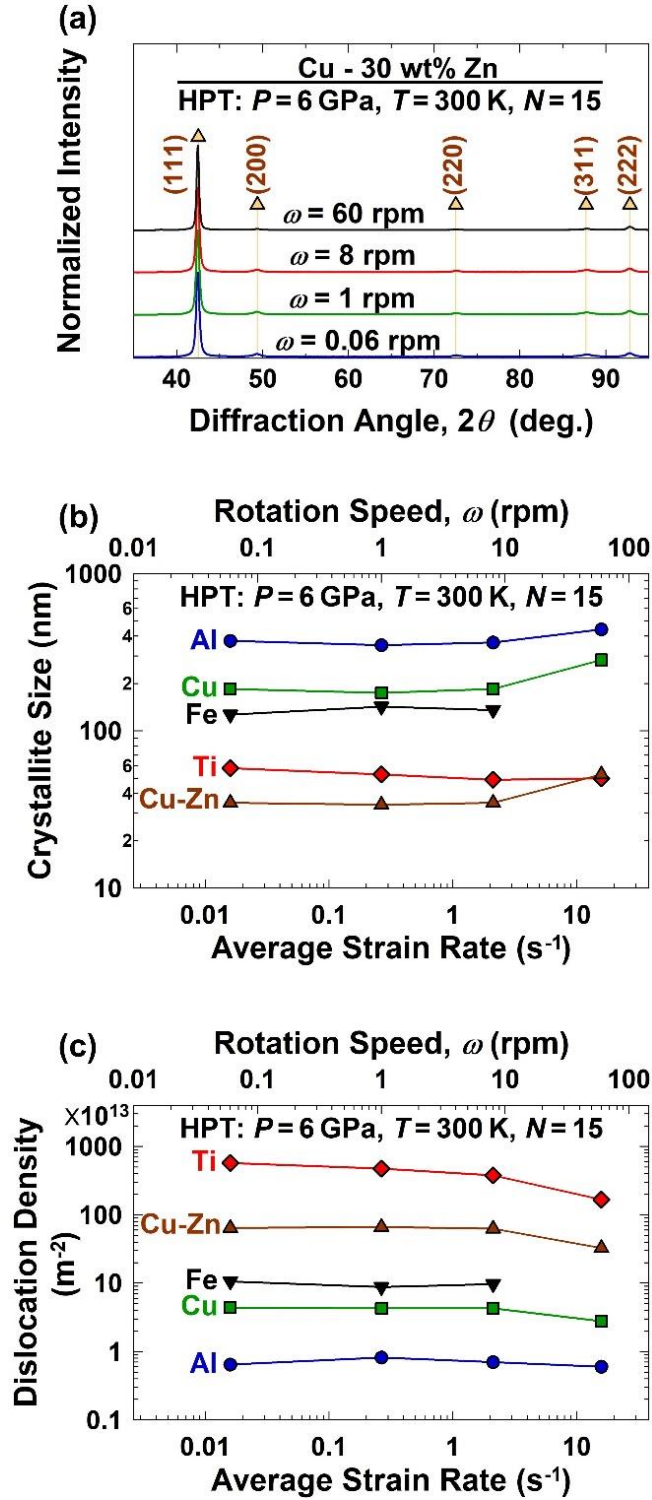

Figure 4. Effect of strain rate on steady-state microstructure. (a) XRD profiles and variations of (b) crystallite size and (c) dislocation density versus average strain rate for (a) Cu-Zn and (b, c) Al, Cu, Cu-Zn, Ti and Fe processed by HPT for 15 rotations with various rotation speeds.

Fig. 5a-d illustrates the TEM bright-field images, dark-field images and SAED patterns for Cu-Zn samples, which were expected to show the most significant changes by strain rate changes due to the solid-solution effect. Fig. 5e summarizes the mean grain sizes versus the average strain rate, in which the average grain sizes were determined from the orthogonal sizes of bright regions in the dark-field images for 20-80 grains. Both bright-field and dark-field images confirm the presence of UFG microstructures in all regions, although the grain boundaries are not well-defined in these micrographs due to significant lattice distortions within the grains. The ring shape of the SAED pattern also suggests the presence of nanograins with the FCC structure. Fig. 5e indicates that the average grain size is independent of the strain rate, although the grain size slightly increases with increasing the rotation speed to 60 rpm in good agreement with the crystallite size measurements in Fig. 4b. The significant reduction of grain size to the nanometer level in Cu-Zn should be due to the interaction of solute atoms and dislocations with increased multiplication rate [10], although some studies attributed this feature to the low stacking fault energy of the alloy [7,8].

Fig. 6 shows the typical steady-state microstructures of (a) Al, (b) Cu, (c) Ti and (d) Fe and Table 1 compares the average crystallite and grain sizes determined in this study with some earlier publications on Al 1050 [17,31], Cu [7-9,32], Cu-Zn [7,8,10,33], Ti [17,34,35] and Fe [9,36,37]. In Table 1, the crystallite sizes are the average of the values at four rotation speeds selected in this study, and grain sizes are the average sizes at a constant rotation speed. It is evident that the measured sizes in this study are reasonably consistent with earlier publications, although crystallite sizes examined by XRD are naturally smaller than the grain sizes measured by TEM.

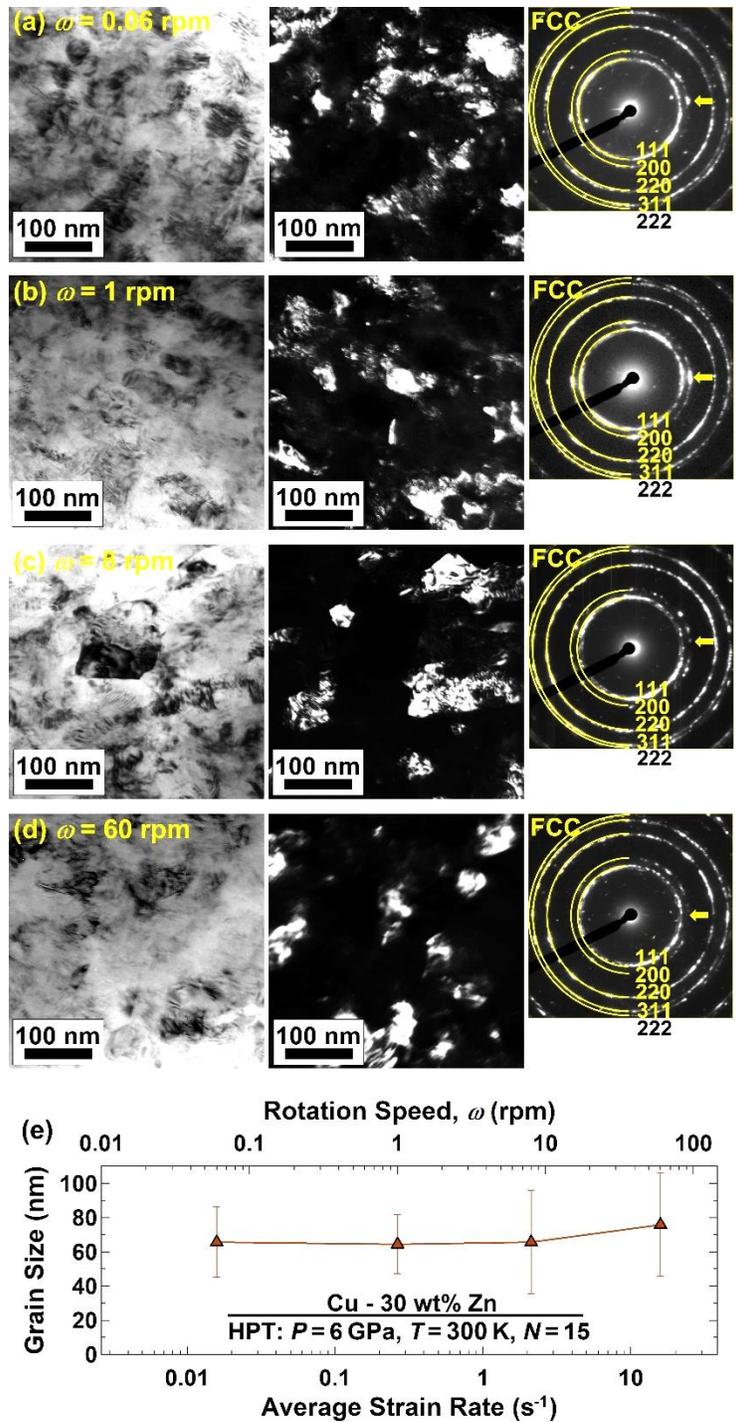

Figure 5. Effect of strain rate on steady-state grain size. (a-d) TEM bright-field images (left), dark-field images (center) and SAED patterns (right) and (e) variation of mean grain size versus average strain rate for Cu-Zn processed by HPT for 15 turns, with rotation speeds of (a) 0.06 rpm, (b) 1 rpm, (c) 8 rpm and (d) 60 rpm. Dark-field images were taken with diffracted beams indicated by arrows in SAED patterns.

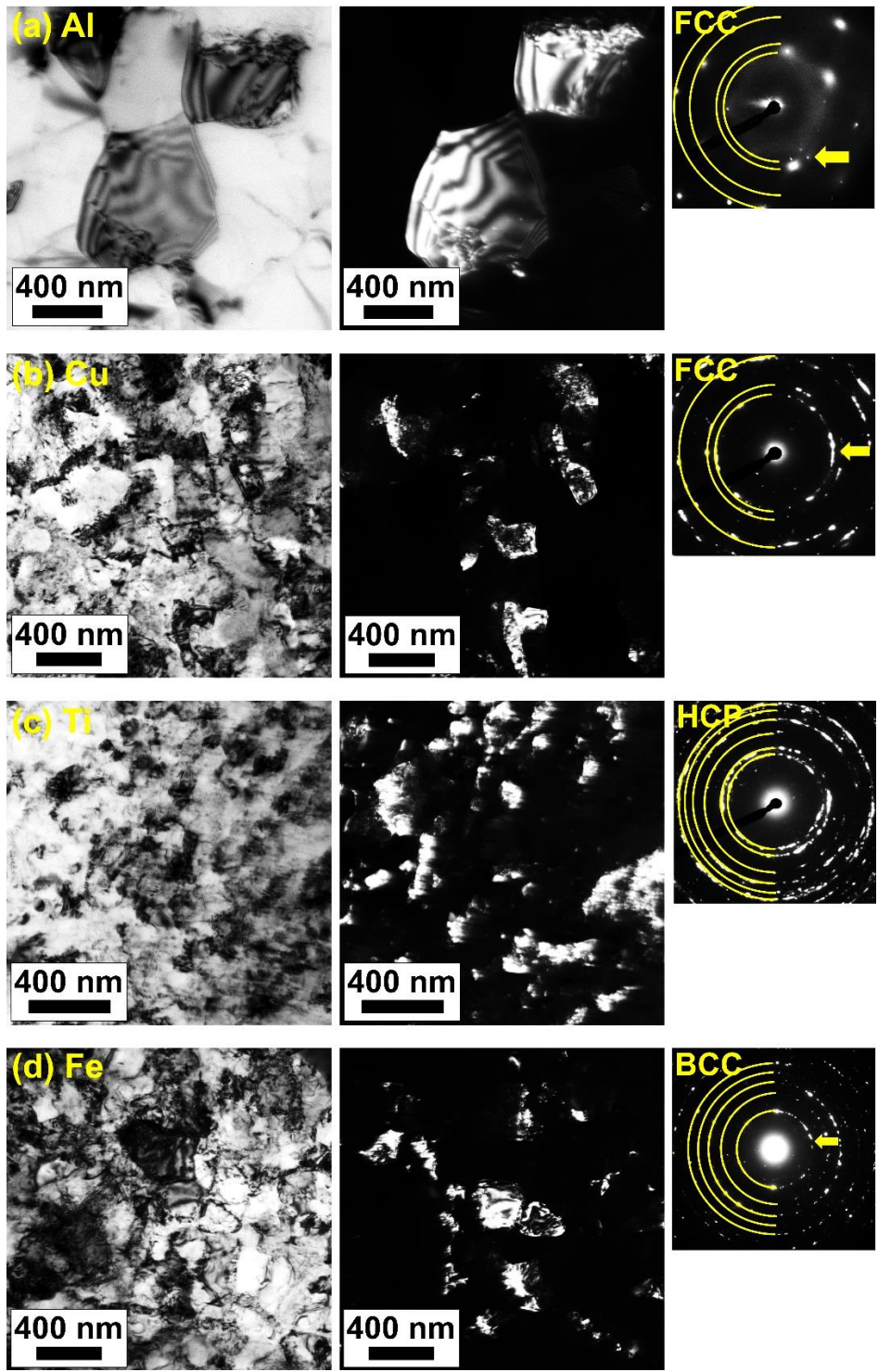

Figure 6. Presence of ultrafine grains in severely deformed pure metals. TEM bright-field images (left), dark-field images (center), and SAED patterns (right) for (a) Al, (b) Cu, (c) Ti and (e) Fe at steady state. Dark-field images were taken with diffracted beams indicated by arrows in SAED patterns.

Table 1. Comparison of crystallite size and grain size measured in this study with those reported in the literature.

| Material | Crystallite Size (nm) | | Grain Size (nm) | |
| --- | --- | --- | --- | --- |
| | This Study | Literature | This Study | Literature |
| Al | 384±40 | --- | 504±226 | 500 [17], 600 [31] |
| Cu | 207±51 | 84 [7], 59 [8] | 273±130 | 200 [9], 290 [32] |
| Cu-Zn | 39±9 | 17 [7], 30 [8] | 64±17 | 74 [33], 75 [10] |
| Ti | 52±4 | 43 [35] | 149±184 | 200 [34], 200 [19] |
| Fe | 135±38 | 87 [37] | 226±119 | 200 [9], 200 [36] |

**Discussion**

Two questions arise from the current study. (i) What are the possible reasons for the independence of steady-state microstructure and flow stress on the strain rate in SPD processing? (ii) What are the reasons for the inconsistency between the conclusion of this study and those from some earlier studies?

Regarding the first question (i), it is well known that a high strain rate and a low processing temperature in metal forming increase the accumulation rate of lattice defects and enhance the fragmentation of grains [3,4]. It was suggested that the effects of strain rate and temperature can be suitably quantified by the Zener-Hollomon parameter ($Z$) not only in low strain levels [4] but also for the steady state [14]. One can understand the steady state in terms of the defect generation rate compared to that of annihilation: with increasing the strain level, the densities of lattice defects reach critical values which launch effects of dynamic recovery, recrystallization and grain boundary migration [3]; this way, a balance between generation and annihilation of lattice defects is reached [3,4]. While the increase of lattice defect densities and/or grain fragmentation leads to strain hardening, the onset of dynamic recovery, recrystallization and grain boundary migration, however, implies strain softening so that in total, no change in overall microstructural features and flow stress occurs [3-5]. An increase in homologous processing temperature reduces the rate of grain fragmentation, enhances the rate of dynamic recrystallization and accordingly leads to an increase in the steady-state crystallite/grain size, as reported in numerous publications [1-6]. An increase in the strain rate enhances the rate of dislocation generation and grain fragmentation but also increases the rate of dynamic recrystallization [21,22]. Unlike the absolute temperature which appears in exponential form in the Zener-Hollomon parameter, the strain rate enters proportionally, and thus the rate of grain fragmentation is less sensitive to the strain rate than to the temperature [4,11]. However, the rate of dynamic recrystallization is also directly proportional to the strain rate [21,22]. Therefore, one may expect that the effect of strain rate on the balance between grain fragmentation and dynamic recrystallization is insignificant, a fact which has - at least within the achievable measurement resolution - been experimentally observed in the current study.

The second question (ii) concerns the contradicting conclusions reported in other studies on the significance of shear strain rate on microstructural evolution during SPD [12-20]. It should be noted that the focus of the current study is on the steady-state microstructure. With our experiments, in order to make sure that the microstructure is really at a steady state, the HPT process was conducted for 15 rotations which correspond to a maximum shear strain of 590. The authors wonder whether the rate dependence reported in some studies may arise from the fact that the microstructural features were not still at a steady state for given deformation modes. Another source of discrepancies between other studies [12-20] and ours may be the fact that the *ex-situ* and *in-situ* measurements of strength do not combine because of static recrystallization effects that may take place after the *ex-situ* experiments, i.e. during the unloading event before the *ex-situ* strength measurements are done [38,39]. Concerning the *in-situ* experiments such as those reported in [14], the way of torque measurement may be different for different facilities used, i.e. imply different contributions of friction which may distort the results to make them strain rate-dependent.

Some other factors may also affect the current experimental observations. One of those is the resolution limit of hardness test, torque measurement, XRD and TEM. Concerning the mechanical measurements, their error is usually within +/- 3%; XRD and TEM are expected to be sensitive enough to reveal visible changes by variation of strain rate from 0.004 to 20 $s^{-1}$, but none of those changes have been observed within the current investigation. The same is true for the earlier deviating studies [12-20] as they did not report significant changes in microstructure at the steady state or close to the steady-state conditions.

Another factor is the temperature rise at high strain rates which can influence the evolution of microstructure, particularly when adiabatic shear bands are formed. However, it was shown using both experimental measurements and finite element modeling that the temperature rise during HPT processing is not so significant because massive anvils connected to large metallic plates act as heat sinks for a small disc sample [29,30]. In this study, the magnitude of temperature rise, which was measured during the process using a thermocouple located 10 mm away from the disc in the upper anvil or by an optical thermometer, was quite small (< 323 K) at least for the rotation speeds of 0.06, 1 and 8 rpm, i.e. strain rates up to 3 $s^{-1}$. Only at the highest speed of 60 rpm applied, some noticeable localized temperature rise may have occurred leading to effects of dynamic and static recrystallization (Fig. 2a, 2b, 4b, 4c). For the samples processed with 60 rpm, the crystallite size slightly increases, and dislocation density slightly decreases (Fig. 4), leading to a slight softening (Fig. 2). However, for the other rotation speeds selected, significant change neither in crystallite size nor in dislocation density occurs and accordingly the hardness/strength is also constant. Taken altogether, the main reason for the independence of the steady state on strain rate appears to be the fairly parallel change in the rates of hardening (crystallite/grain fragmentation) and those of the softening (dynamic recrystallization) phenomena.

**Conclusions**

In summary, the current study on the processing of four metals with different melting points (Al, Cu, Ti and Fe) and a Cu-Zn alloy using HPT with strain rates of 0.004 to 20 $s^{-1}$ confirms that the steady-state microstructure, hardness and shear stress are independent of strain rate, at least within the resolution limits of XRD, TEM, torque measurement and microhardness tests. These findings suggest that although a high strain rate is effective in the enhancement of crystallite/grain fragmentation and defect accumulation at the early stages of straining, the final microstructure at the steady state, which is achieved by a balance between crystallite/grain fragmentation and dynamic recrystallization, is reasonably independent of strain rate.

**Acknowledgments**

This work has been supported in part by the Light Metals Educational Foundation of Japan, in part by Grants-in-Aid for Scientific Research on Innovative Areas (JP19H05176 & JP21H00150) and Challenging Research (Exploratory) (JP22K18737) from the MEXT, Japan, and in part by the "Metals and Alloys under Extreme Impacts" Laboratory of Eurasian Center of Excellence, USATU (assignment #075-03-2021-014/4).